# Raman Scattering near Metal Nanostructures

M. Scalora[1], M. A. Vincenti[2], D. de Ceglia[2], M. Grande[3], J. W. Haus[4]

[1] *Charles M. Bowden Research Center RDMR-WSS, RDECOM, Redstone Arsenal, Alabama 35898-5000, USA*

[2] *AEgis Technologies Group, 410 Jan Davis Dr., Huntsville, AL 35806*

[3] *Dipartimento di Elettrotecnica ed Elettronica, Politecnico di Bari, Via Re David 200, 70125 Bari, Italy*

[4] *Electro-Optics Program, University of Dayton, 300 College Park, Dayton, OH 45469*

## ABSTRACT

We study Raman scattering in active media placed in proximity of different types of metal nanostructures, at wavelengths that display either Fabry-Perot or plasmonic resonances, or a combination of both. We use a semi-classical approach to derive equations of motion for Stokes and anti-Stokes fields that arise from quantum fluctuations. Our calculations suggest that local field enhancement yields Stokes and anti-Stokes conversion efficiencies between five and seven orders of magnitudes larger compared to cases without the metal nanostructure. We also show that to first order in the linear susceptibility the local field correction induces a dynamic, intensity-dependent frequency detuning that at high intensities tends to quench Raman gain.

**Introduction**

Stimulated Raman scattering (SRS) [1, 2] involves the inelastic scattering of light and as such it is a ubiquitous process. It occurs under a variety of circumstances, conditions, and in a vast number of materials, including air. The interaction usually leads to the generation of red-shifted and blue-shifted light from quantum fluctuations. If a medium is illuminated by a pump field, and if the active cell is sufficiently long, then the magnitude of the generated fields can reach macroscopic values. The system is therefore ideal to study macroscopic manifestations of quantum noise, such as energy [3] and beam pointing fluctuations [4, 5, 6], and the Raman soliton [7], for example. These fluctuations arise from the uncertainty associated with the system's initial conditions, so that both energy and direction of propagation differ every time an experiment is performed. On the other hand, the Raman soliton comes as a result of a π-phase shift in the fluctuating field, which temporarily inverts the sign of gain pushing energy back towards the pump, giving rise to bright (pump) and dark (Stokes) solitary waves.

In surface enhanced Raman scattering (SERS) one usually observes high-gain amplification of spontaneously generated Raman signals from molecules that are placed near metallic structures. Fleischmann, Hendra, and McQuillan are credited with the original observation of SERS [8], which motivated many publications on what for all practical purposes may be referred to as a new field of study [9-34, for example]. Although the main idea advanced in most of these works is to exploit local field enhancement due to the formation of plasmonic features and surface plasmons [23], it is also known that the band structure of metal gratings that contain aperture arrays also display Fabry-Perot resonances where field localization near cavity edges and corners may surpass that achievable at plasmonic resonances [35, 36]. Therefore, the enhancement of linear and nonlinear optical phenomena in metal gratings is multifaceted and not due exclusively to plasmonic resonances [37].

Another aspect of SERS regards stimulated processes, which for example were reported from the surface of a 5nm copper phtalocyanine evaporated on top of a silver grating [38]. More recent studies have led to the implementation of surface-enhanced, femtosecond stimulated Raman spectroscopy (SE-FSRS) [39-41]. In reference [39] active molecules were embedded in environments containing gold nano-antennas consisting of two metal spheres, with reported SERS enhancement factors in the range of $10^4$-$10^6$ compared to single molecule excitation. Time-resolved SE-FSRS can be used as an effective tool to study the unique, ultrafast dynamics connected to plasmonic materials. We note that an extensive analysis of both SERS and fluorescence near isolated nanoparticles has already been carried out [42]. In view of the many contributions to the subject matter, our aim here is certainly not to reinvent the wheel. However, we take an approach usually absent in most descriptions, designed to include novel effects using practical equations of motion that may be used under a variety of circumstances and conditions, in systems with a degree of complication that analytical solutions are not achievable.

A few configurations that we will consider in our study of surface enhanced processes are shown in Fig.1, where we illustrate the elementary cells of several metal-based nanostructures surrounded or filled by a Raman active medium. The aperture system depicted in Fig.1(a) was studied in details in reference [35]. For deeply sub-wavelength apertures transmittance of TE-polarized waves (E-field polarized along the *x*-direction) is negligible. A plasmonic band gap appears for TM-polarized light (E-field in the plane of incidence *y-z*) in ranges where the periodicity *p* of the grating matches the effective wavelength of the surface plasmon supported at the metal interface. Fabry-Perot resonances emerge at multiples of layer thickness $w=\lambda/2n_{eff}$ [35], where $n_{eff}$ is the effective index of the slit waveguide.

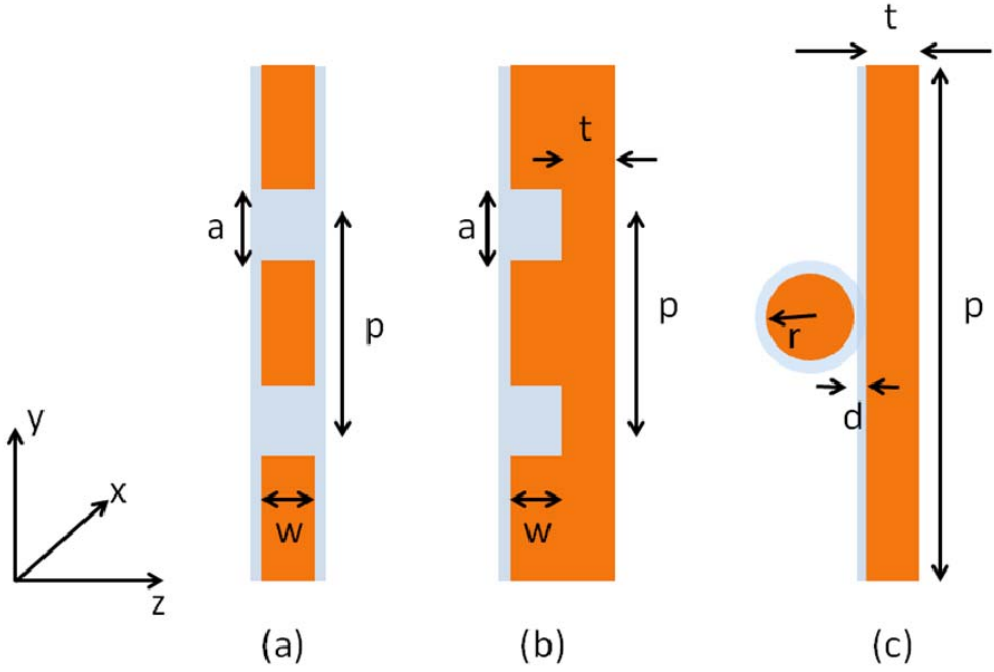

**Fig.1** (a) Metal grating that contains aperture arrays surrounded by a Raman active medium (light blue area) that also fills the slits. (b) Metal grating that contains grooves only, surrounded by Raman active medium. (c) Unit cell of a 1D periodic array composed of metal nanowires placed near a metal substrate.

The second possible configuration is shown in Fig.1(b) and consists of a thick metal grating that contains only grooves. In this scenario the incident light is mostly reflected and one expects dips in the reflection spectrum that represent energy transferred to the grating in the form of surface waves. Under realistic conditions, which also means that metal corners are slightly rounded with a radius of curvature of a few nanometers to avoid singular field behavior, in both Figs.1(a) and (b) the electric field intensity $|\mathbf{E}|^2$ is amplified between 600 and 1200 times at the corners of the cavity and at the edges of the groove, clearly offering new opportunities for nonlinear optical interactions [36].

As another example of a metal-based grating we consider metal nanowires placed near a metal substrate a few tens of nanometers thick, depicted in Fig.1(c). This geometry [43] is of interest for nonlinear optical applications for at least two reasons: (i) the field tends to become singular as the nanowire approaches the substrate; (ii) broadband absorption peaks appear with maxima that reach nearly 100%. These absorption values suggest these structures may be used for applications to energy harvesting. Of course, using different metals the same geometries we have explored work well in different wavelength ranges with similar results. We will discuss the Raman emission properties of the structures outlined in Fig.1 after we describe the model in the following section.

**The Model**

We now derive a semiclassical model of Raman scattering from active molecules placed near generic nanostructures similar to those depicted in Fig.1, valid in both spontaneous and stimulated regimes. By semiclassical here we mean that the fields and the dielectric function of the metal or other material composing the grating are treated classically, while the Raman active molecules are described quantum mechanically using a density matrix approach. In the range 300-1200nm the dielectric function of the metal is adequately described by a combined Drude-Lorentz model as follows [36]:

$$\ddot{\mathbf{P}}_f + \tilde{\gamma}_f \dot{\mathbf{P}}_f = \frac{n_{0,f} e^2}{m_f^*} \left(\frac{\lambda_0}{c}\right)^2 \mathbf{E} + \frac{5}{3} \frac{E_F}{m_f^* c^2} \nabla(\nabla \bullet \mathbf{P}_f), \tag{1}$$

$$\ddot{\mathbf{P}}_1 + \tilde{\gamma}_{01} \dot{\mathbf{P}}_1 + \tilde{\omega}_{0,1}^2 \mathbf{P}_1 = \frac{n_{0,1} e^2}{m_b^*} \mathbf{E}, \tag{2}$$

$$\ddot{\mathbf{P}}_2 + \tilde{\gamma}_{02} \dot{\mathbf{P}}_1 + \tilde{\omega}_{0,2}^2 \mathbf{P}_2 = \frac{n_{0,2} e^2}{m_b^*} \mathbf{E}, \tag{3}$$

where $\tilde{\omega}_{p,f}, \tilde{\gamma}_f$ are the scaled plasma frequency and damping coefficient for free electrons, $\tilde{\omega}_{p,1}, \tilde{\omega}_{0,1}, \tilde{\gamma}_{0,1}$ and $\tilde{\omega}_{p,2}, \tilde{\omega}_{0,2}, \tilde{\gamma}_{0,2}$ are the scaled plasma frequencies, resonance frequencies, and damping coefficients for two species of bound electrons; $n_{0,f}$, $n_{0,1}, n_{0,2}$ are free and bound electron densities; and $m_b^*$ are the effective electron masses in the conduction and valence bands of the metal, respectively $m_f^*$ and $m_b^*$ are the effective electron masses in the conduction and valence bands of the metal, respectively. For simplicity we assume $m_f^* = m_b^* = m_e$, where $m_e$ is the free-electron mass. $\mathbf{P}_f, \mathbf{P}_1$, and $\mathbf{P}_2$ are the linear polarizations associated with free-electrons and two separate bound electron species, respectively. The inclusion of bound electrons in the theoretical description of the metal enables a more accurate study of surface processes in disparate wavelength ranges under realistic conditions, as we include the dynamics of inner-core electrons to the dielectric response. The scaled, dimensionless coordinates that appear in Eqs.(1-3) in the form of time derivatives and $\nabla = \frac{\partial}{\partial \tilde{y}} \mathbf{j} + \frac{\partial}{\partial \xi} \mathbf{k}$ are: $\xi = z/\lambda_0, \tilde{y} = y/\lambda_0$, and $\tau = ct/\lambda_0$, and we have chosen a reference wavelength $\lambda_0 = 1 \mu m$. The second term on the right hand side of Eq.(1) is also new in this context, and contains the Fermi energy $E_f$ of the free electron gas as part of an overall nonlocal term that imparts wave-vector dependence to the linear dielectric constant. Under the present circumstances this term can: (i) shift the band structures for each of the nanostructures shown in Fig.1 [36]; (ii) influence linear absorption; and (iii) modify nonlinear conversion efficiencies, depending on the type of metal and surface features. However, it does not change the basic, qualitative aspects of our findings.

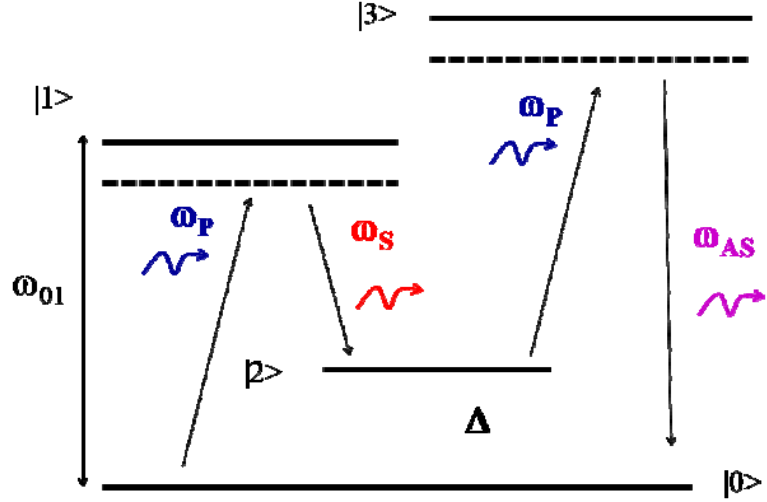

**Fig.2:** Energy level diagram for the generation of Stokes (red-shifted) and anti-Stokes (blue-shifted) photons of frequencies $\omega_S$ and $\omega_{AS}$, respectively, when photons of frequency $\omega_P$ excite the atom. The dashed lines are virtual states located far below levels |1> and |3>. A pump photon excites the system to level |2> producing a Stokes photon, while a second pump photon brings it back down to the ground state via the virtual state below state |3> generating anti-Stokes light.

Adopting the notation and the theoretical development in reference [7], we assume the system consists of a collection of four-level atoms, as depicted in Fig.2. In the dipole approximation, the Hamiltonian (divided by Plank's constant) for this system may be written as:

$$H = \omega_{01}|1><1| + \omega_{03}|3><3| + \Delta|2><2| - \frac{\mu \cdot E}{\hbar}(|1><0| + |0><1| + |1><2| + |2><1| \\ + |3><0| + |0><3| + |3><2| + |2><3|)$$
(4)

where the terms proportional to $|1><1|$, $|2><2|$, and $|3><3|$ are the energies for states $|1>$, $|2>$ and $|3>$, respectively, $|1><0|$ and $|0><1|$ are dipole matrix elements that enable electronic transitions between states |0> and |1>; the remaining terms in the Hamiltonian are similarly described. With reference to the coordinate axes shown in Fig.1, the wave propagates and scatters on the *y-z* plane. We assume the grating is infinite in extent along the *x*-direction, so that the fields are independent of *x*, and refer to *z* as the longitudinal coordinate. Then, a TM-polarized field has E-field components on the *y-z* plane; a TE-polarized field has corresponding H-field components on the *y-z* plane. An incident TM-polarized field may be written as a superposition of all fields present, namely:

$$\mathbf{E} = E_y \mathbf{j} + E_z \mathbf{k} = \left( \sum_l \mathcal{E}_{l,y}(\mathbf{r},t) e^{-i\omega_l t} + \mathcal{E}^*_{l,y}(\mathbf{r},t) e^{i\omega_l t} \right) \mathbf{j} + \left( \sum_l \mathcal{E}_{l,z}(\mathbf{r},t) e^{-i\omega_l t} + \mathcal{E}^*_{l,z}(\mathbf{r},t) e^{i\omega_l t} \right) \mathbf{k}$$

$$\mathbf{H} = H_x \mathbf{i} = \left( \sum_l \mathcal{H}_{l,x}(\mathbf{r},t) e^{-i\omega_l t} + \mathcal{H}^*_{l,x}(\mathbf{r},t) e^{i\omega_l t} \right) \mathbf{i}$$

(5)

where $\mathcal{E}_{l,y}(\mathbf{r},t), \mathcal{E}_{l,z}(\mathbf{r},t)$ and $\mathcal{H}_{l,x}(\mathbf{r},t)$ are generic envelope functions, and $\omega_l$ are carrier frequencies for the $l^{th}$ field. We do not make the slowly varying envelope approximation: in the presence of highly reflective surfaces the envelope functions are allowed to vary rapidly in space and time. In the Heisenberg picture, the wave function may be expanded as a linear superposition of basis states with time dependent probability amplitudes, namely:

$$|\psi(t)> = C_0(t)|0> + C_1(t)|1> + C_2(t)|2> + C_3(t)|3> \quad , \tag{6}$$

where $C_i(t) = <i|\psi(t)>$. The equations of motion for the density matrix elements are:

$$i\dot{\rho}_{ij} = \frac{1}{\hbar}[\rho_{ij}, H], \tag{7}$$

where the matrix elements are defined as:

$$\rho_{ij} = <\psi(t)|i><j|\psi(t)> \quad . \tag{8}$$

The commutator Eq.(7) becomes:

$$i\dot{\rho}_{ij} = \frac{1}{\hbar}\left( <\psi(t)|i><j|H|\psi(t)> - <\psi(t)|H|i><j|\psi(t)> \right). \tag{9}$$

From Eq.(4), the dipole moment operator in this subspace is given by:

$$\boldsymbol{\mu} = \mu_0 \hat{\boldsymbol{\mu}} \left( |1><0| + |0><1| + |1><2| + |2><1| + |3><0| + |0><3| + |3><2| + |2><3| \right). \tag{10}$$

where $\hat{\boldsymbol{\mu}}$ is a unit vector on the y-z plane. The dipole moment's expectation value may be calculated as:

$$<\psi|\boldsymbol{\mu}|\psi> = \mu_0 \hat{\boldsymbol{\mu}} \left( \rho_{10} + \rho_{01} + \rho_{12} + \rho_{21} + \rho_{30} + \rho_{03} + \rho_{32} + \rho_{23} \right), \tag{11}$$

which immediately yields the polarization as:

$$\mathbf{P} = N\mu_0 \hat{\boldsymbol{\mu}} \left( \rho_{10} + \rho_{01} + \rho_{12} + \rho_{21} + \rho_{30} + \rho_{03} + \rho_{32} + \rho_{23} \right), \tag{12}$$

where N is the molecular density. The equations of motion for the matrix elements (without relaxation terms) are then developed using Eq.(9) for each Cartesian coordinate, $k = x, y, z$, so that:

$$i\dot{\rho}^k_{11} = -\frac{\mu_k E_k}{\hbar} \left( \rho^k_{10} - \rho^k_{01} + \rho^k_{12} - \rho^k_{21} \right) \quad , \tag{13a}$$

$$i\dot{\rho}^k_{22} = \frac{\mu_k E_k}{\hbar} \left( \rho^k_{12} + \rho^k_{32} - \rho^k_{21} - \rho^k_{23} \right) \quad , \tag{13b}$$

$$i\dot{\rho}_{33}^k = -\frac{\mu_k E_k}{\hbar}\left(\rho_{03}^k + \rho_{23}^k - \rho_{30}^k - \rho_{32}^k\right) \quad , \tag{13b}$$

$$i\dot{\rho}_{00}^k = \frac{\mu_k E_k}{\hbar}\left(\rho_{10}^k - \rho_{01}^k + \rho_{30}^k - \rho_{03}^k\right) \quad , \tag{13c}$$

$$i\dot{\rho}_{01}^k = \omega_{01}\rho_{01}^k + \frac{\mu_k E_k}{\hbar}\left(\rho_{11}^k - \rho_{00}^k + \rho_{31}^k - \rho_{02}^k\right), \tag{13d}$$

$$i\dot{\rho}_{12}^k = (\Delta - \omega_{01})\rho_{12}^k + \frac{\mu_k E_k}{\hbar}\left(\rho_{02}^k + \rho_{22}^k - \rho_{11}^k - \rho_{13}^k\right) \quad , \tag{13e}$$

$$i\dot{\rho}_{02}^k = \Delta\rho_{02}^k + \frac{\mu_k E_k}{\hbar}\left(\rho_{12}^k - \rho_{01}^k + \rho_{23}^k - \rho_{03}^k\right) \quad , \tag{13f}$$

$$i\dot{\rho}_{03}^k = \omega_{03}\rho_{03}^k + \frac{\mu_k E_k}{\hbar}\left(\rho_{13}^k + \rho_{33}^k - \rho_{00}^k - \rho_{02}^k\right) \quad , \tag{13g}$$

$$i\dot{\rho}_{32}^k = (\omega_{03} - \Delta)\rho_{32}^k + \frac{\mu_k E_k}{\hbar}\left(\rho_{31}^k + \rho_{33}^k - \rho_{02}^k - \rho_{22}^k\right) \quad . \tag{13h}$$

Eqs.(13) may be simplified by assuming that $\omega_{01}$, $\omega_{03}$, and $\omega_{23}$ are much larger than any other frequency in the problem (off-resonant Raman), i.e. states |1> and |3> remain unpopulated. It follows that $\rho_{11}^k = \rho_{33}^k \approx 0$, $\rho_{03}^k \approx \rho_{01}^k \approx \frac{\mu_k E_k}{\hbar}\left(\rho_{00}^k + \rho_{02}^k\right)$, $\rho_{32}^k \approx \rho_{12}^k \approx \frac{\mu_k E_k}{\hbar}\left(\rho_{22}^k + \rho_{02}^k\right)$, and conservation of probability takes the form $\rho_{22}^k + \rho_{00}^k = 1$. By further defining the function $Q_k \equiv -ie^{i\delta t}\rho_{02}^k$, the population inversion $W_k \equiv \rho_{22}^k - \rho_{00}^k$, and the Rabi frequency $\Omega_k \equiv \frac{1}{\omega_{01}}\left(\frac{\mu_k E_k}{\hbar}\right)^2$, we may write equations of motion for a reduced system, equivalent to those of a two-level atom, as follows [7]:

$$\dot{Q}_k = -i(\Delta - \delta)Q_k - \Omega_k W_k e^{i\delta t} \quad , \tag{14}$$

$$\dot{W}_k = 2\Omega_k\left[Q_k e^{-i\delta t} + Q_k^* e^{i\delta t}\right] \quad . \tag{15}$$

The detuning $\delta \equiv \omega_l - \omega_{l+1} > 0$; the frequencies are such that if $l$ represents the pump field, then $l+1$ represents a field of lower frequency, i.e. the Stokes field, with the further simplification that $\delta$ remain constant for any field pair. Choosing $\boldsymbol{\mu} = \mu_y\mathbf{j} + \mu_z\mathbf{k}$ allows coupling to both transverse and longitudinal electric field components. The transverse field can play a major role away from plasmonic resonances. For simplicity we may choose $\mu_y = \mu_z = \mu_0$. Substituting Eq.(5) into Eqs.(14-15) we have:

$$\dot{Q}_k = -i(\Delta - \delta)Q_k - \frac{\mu_0^2}{2\hbar^2 \omega_{01}} \left( \sum_l \mathcal{E}_{l,k}(\mathbf{r},t)\mathcal{E}_{l+1,k}^*(\mathbf{r},t) \right) W_k \qquad ,(16)$$

$$\dot{W}_k = \frac{\mu_0^2}{\hbar^2 \omega_{01}} \left( Q_k^* \sum_l \mathcal{E}_{l,k}(\mathbf{r},t)\mathcal{E}_{l+1,k}^*(\mathbf{r},t) + c.c. \right) \qquad . \qquad (17)$$

The final assumptions that complete the model consist of: (i) a stipulation that the total number of active molecules is always much greater than the number of photons, so that the system can be said to be in the ground state at all times, i.e. $W_k \approx -1$. This assumption may breakdown within high-intensity hot spots, where field intensities may be enhanced by many orders of magnitude, so that local photon density could overwhelm the molecular density leading to saturation phenomena and/or breakdown, perhaps causing stimulated emissions to overtake spontaneous processes. However, we presently continue to assume the condition $W_k \approx -1$ holds. (ii) The introduction of a phenomenological, non-Hamiltonian, collisional dephasing term that causes the coherence $Q_k$ to decay, along with a corresponding quantum noise source term [44]. After scaling the coordinates as in Eqs.(1-3), Eqs.(16-17) reduce to a single equation, namely:

$$\dot{Q}_k = \left(-\tilde{\gamma} - i\tilde{\Delta}\right)Q_k + \frac{\mu_0^2 \lambda_0}{2\hbar^2 \omega_{01} c} \left( \sum_l \mathcal{E}_{l,k}(\mathbf{r},\tau)\mathcal{E}_{l+1,k}^*(\mathbf{r},\tau) \right) + F_k(\mathbf{r},\tau). \qquad (18)$$

The new scaled coefficients are $\tilde{\gamma} = \gamma \frac{\lambda_0}{c}$ and $\tilde{\Delta} = (\Delta - \delta)\frac{\lambda_0}{c}$. $F_k(\mathbf{r},\tau)$ is a Gaussian, white noise source term having variance $2\tilde{\gamma}/N_0$, where $N_0$ is now the number of molecules, and such that $<F_k(\mathbf{r},\tau)> = 0$ and $<F_k(\mathbf{r},\tau)F_k(\mathbf{r}',\tau')> = 0$. The relative amplitudes of the field coupling coefficient $\frac{\mu_0^2 \lambda_0}{2\hbar^2 \omega_{01} c}$ and the amplitude of the fluctuation term $F_k(\mathbf{r},\tau)$ in Eq.(18) determine the relative amounts of generated spontaneous and stimulated emissions. Our solution then proceeds within the context of the stochastic model of Raman scattering [45-47]. Briefly, in this model the ensemble averages are obtained by performing many calculations with different, random initial conditions only at t=0, rather than introducing Gaussian white noise at every point in space for each time step. The interested reader will find more details about the model in references [45-47]. Using the definitions and results above, the Cartesian polarization components resulting from Eq.(12) are:

$$P_k = \chi_L E_k + i\chi_L \left( Q_k E_k e^{-i\delta t} - Q_k^* E_k e^{i\delta t} \right) \qquad , \qquad (19)$$

where $E_k = E_k(\mathbf{r},\tau)$ is the total, real field of Eq.(5) along the *k*-direction, and $\chi_L = \dfrac{4N\mu_0^2}{\hbar\omega_{01}}$ is the linear, background susceptibility of the Raman active medium. Finally, substituting Eq.(5) into Eq.(19) and retaining only nonlinear polarization terms that oscillate at the pump, Stokes and anti-Stokes frequencies we have [7]:

$$\begin{aligned}
\mathcal{P}^{NL}_{\omega_p,k} &= i\chi_L\left[Q_k \mathcal{E}_{S,k} - Q_k^* \mathcal{E}_{AS,k}\right] \\
\mathcal{P}^{NL}_{\omega_S,k} &= -i\chi_L Q_k^* \mathcal{E}_{P,k} \\
\mathcal{P}^{NL}_{\omega_{AS},k} &= i\chi_L Q_k \mathcal{E}_{P,k}
\end{aligned} \qquad (20)$$

where $\mathcal{E}_{P,k}$, $\mathcal{E}_{S,k}$, $\mathcal{E}_{AS,k}$ are the generic pump, Stokes, and anti-Stokes field envelopes, respectively, used in Eq.(5). The total polarization is evaluated as a linear superposition of the polarizations in Eqs.(1-3) and Eqs.(20), which are then combined and inserted back into Maxwell's equations. Retention of the nonlinear pump source term in Eq.(20) generalizes the dynamics beyond the undepleted pump approximation. In summary, Eqs.(1-3) describe the metal grating, while Eqs.(20) portray the Raman active medium that surrounds the different gratings shown in Fig.1. For simplicity, we choose $\mu_0 = ea$, where *e* is the electron charge and *a* is the Bohr radius, $\omega_{01} \approx \omega_{03} \approx \omega_{23} = 10^{15}/\text{sec}$, and the molecular density *N* is chosen so that the linear background index of the Raman active medium is $n_L \approx 1.001$. Although this corresponds to dry air ($N_{\text{dry air,STP}} \approx 2.5\times 10^{19}/cm^3$) under pressure, the nature of the Raman active medium is arbitrary and the parameters could just as easily reflect a solid layer of material with complex index of refraction [48]. For spontaneous emission typical incident pulse durations may be chosen so that one is able to resolve the characteristic spectral features of each structure, and will vary from a few tens to a few hundred femtoseconds. As a result collisions, which in gases matter usually for nanosecond time scales, have little impact on field dynamics. Finally, a detuning may be introduced and even allow the use of seed Stokes and/or anti-Stokes fields to study stimulated processes, but presently it makes little difference to the overall qualitative aspect of our results. The material equations of motion are solved in the time domain together with Maxwell's equations using spectral methods based on a formulation of the extended split-step, fast Fourier transform algorithm [36].

**Raman Enhancement factor**

In most of the available literature an assessment of Raman gain for any given structure is defined via the evaluation of the Stokes gain or enhancement factor:

$$G = \frac{<|E_{\omega_p}^{local}|^2>}{<|E_{\omega_p}^{inc}|^2>} \frac{<|E_{\omega_S}^{rad}|^2>}{<|E_{\omega_p}^{inc}|^2>}, \qquad (21)$$

where the brackets denote an integral over either an active surface or a line, depending on the structure' symmetry. This expression is a simple appraisal of the product of the effective density of states of the pump and one paired field. In fact, a similar expression may be derived from simplified equations of motion only for a single field pair. $G$ is expressed as the product of the local ($\sim \frac{<|E_{\omega_p}^{local}|^2>}{<|E_{\omega_p}^{inc}|^2>}$) and radiation ($\sim \frac{<|E_{\omega_s}^{rad}|^2>}{<|E_{\omega}^{inc}|^2>}$) intensity enhancement factors for the pump and Raman frequencies, $\omega_p$, $\omega_S$, and/or $\omega_{AS}$, respectively, normalized with respect to the incident pump field intensity, $|E_{\omega_p}^{inc}|^2$ [48]. In most studies, $G$ is simplified even further by assuming zero Stokes (or anti-Stokes) shift, so that:

$$G_s = <|E_{\omega_p}^{local}|^4> / <|E_{\omega_p}^{inc}|^4>. \qquad (22)$$

One of the difficulties associated with $G$ and $G_s$ is that they are evaluated using linear field profiles obtained by pumping the system at the frequencies of interest in the absence of an active medium. Instead, the gain factor should be calculated by solving equations of motion for atoms that: (i) are distributed inside a cavity environment; (ii) are driven by a pump tuned to a different wavelength; and (iii) may possibly become inverted within local hot spots with repercussions on the dynamics. This means that unless the structure is non-resonant, or has a fairly simple topology, as in an ordinary Fabry-Perot cavity, one should not nurture the expectation that a simple linear analysis of a complex nanostructure will always yield reliable results in the nonlinear regime. For gratings that display strongly modulated plasmonic band structures pump, Stokes and anti-Stokes frequencies can exhibit very different field localization properties relative to each other and to the pump. These circumstances are similar to what occurs in photonic band gap structures of finite length, where field localization properties and phase matching conditions must be sought carefully near the band edge in order to maximize field overlap and harmonic generation [50]. Likewise, the spatial distribution of dipoles inside a planar nanocavity must be chosen to match closely local electric field intensity maxima for efficient stimulated or spontaneous emission, otherwise no enhancement occurs even though the density of modes is high [51, 52]. In SRS, effects due to phase matching conditions are well-known for large gain lengths (centimeter or meter size Raman cells) when at least one Stokes and one anti-Stokes fields are present, while nothing is known for the enormous local field intensities and short gain lengths that may characterize plasmonic hot spots. In the former case, Stokes and anti-Stokes fields can influence each other's conversion efficiencies as a function of wave vector detuning, $\delta k = k_{AS} + k_S - 2k_P$, in a counterintuitive way so that both signals are suppressed as $\delta k \Rightarrow 0$ [2, 53]. All these issues are brushed aside in the implementation of Eqs.(21-22). In the context of SERS, concerns about the predictive utility of $G_S$ were raised in reference [54], for example, where corrections terms were derived to improve upon the perceived shortcomings of the postulated $<|E_{\omega_p}^{local}|^4>$ dependence. An accurate evaluation of the radiation enhancement factor thus requires *a priori* knowledge of the solution of the problem. As a result, $G_s$ may not be considered an accurate predictive measure of gain if the photonic or plasmonic band structure is strongly modulated, as is often the case, while both $G$ and $G_S$ can be poor predictors of relative gain even for the case of zero Stokes shift, at least according to our model.

**Aperture Grating**

Let us begin our discussion by investigating the Raman signals emitted from the aperture grating shown in Fig.1(a), where the metal is silver. In Fig.3 we display the *linear* transmission and absorption spectra for the grating and three possible tuning arrangements. For configuration ($AS_1,P_1,S_1$) the fields are tuned near the same resonance, and display similar spatial profiles and good overlap. Arrangements ($AS_2,P_2,S_2$) and ($AS_3,P_3,S_3$) describe fields tuned across large wavelength ranges and their spatial overlap is uncertain. In Fig.4 we show the spatial distributions of the electric field intensities at locations that in Fig.3 denote pump tuning, including 1115nm, 625nm and 575nm. The three results in Fig.4 also provide insight into the profiles of the generated fields and the degree of overlap with the pump when they are tuned to those wavelengths. One peculiarity of the Fabry-Perot resonance (Fig.4(a)) is that even though its cavity quality factor $Q$ is misleadingly small ($Q \equiv \lambda/\delta\lambda \sim 4$), near the corners of the cavity the total electric field intensity may be amplified 600-700 times relative to the incident field intensity. All things being equal, this local field intensity surpasses by nearly one order of magnitude the intensity found at the plasmonic resonance (left of the gap, Fig.4(c), where in contrast $Q \sim 190$) [35]. For this type of grating pumping narrow plasmonic resonances (575nm) may not yield the best nonlinear results. Rectangular nano-apertures like those in Fig.1 [35] behave differently compared to either single holes [55] or nano-arrays [56], where the enhancement of SERS has been observed. Sub-wavelength rectangular apertures are dominated by the fundamental, TM-polarized, waveguide (Fabry-Perot, or FP) modes propagating in the longitudinal direction. In contrast, circular holes do not support TEM modes.

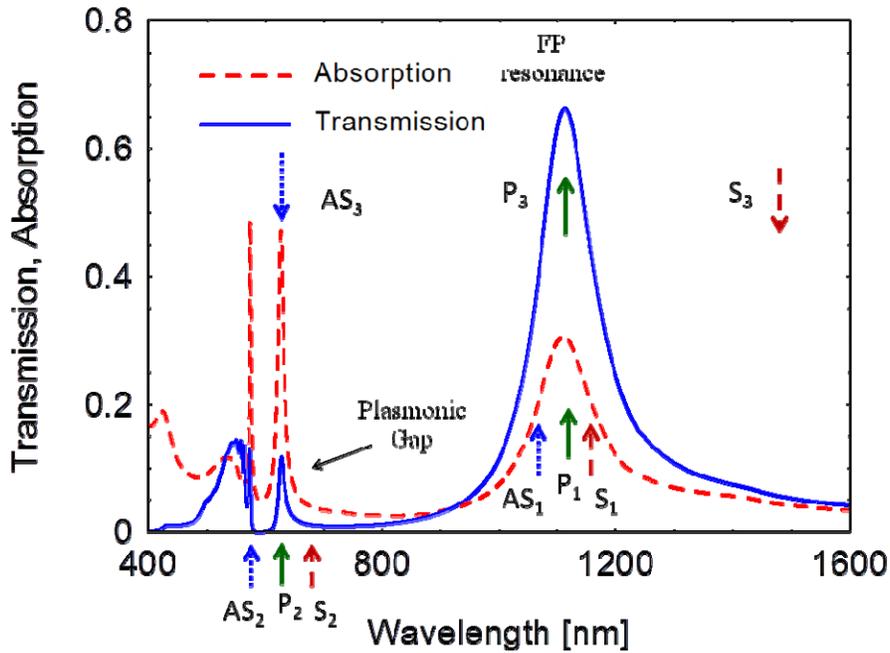

**Fig.3:** Transmission and absorption vs. wavelength for the silver grating in Fig.1(a), for $a$=32nm, $w$=300nm and $p$=566nm. A plasmonic band gap appears when $\lambda_{sp} \sim p$, i.e. $\lambda_{sp}$=578nm. A broad, Fabry-Perot (FP) resonance is centered at ~1115nm. The solid green, dashed red and dotted blue arrows indicate pump, Stokes, and anti-Stokes tuning, respectively, for three configurations. The model allows the fields to be tuned across a wide wavelength range that includes the FP resonance and the plasmonic band gap. The numbered groups (AS,P,S) stand for anti-Stokes, pump, and Stokes, respectively.

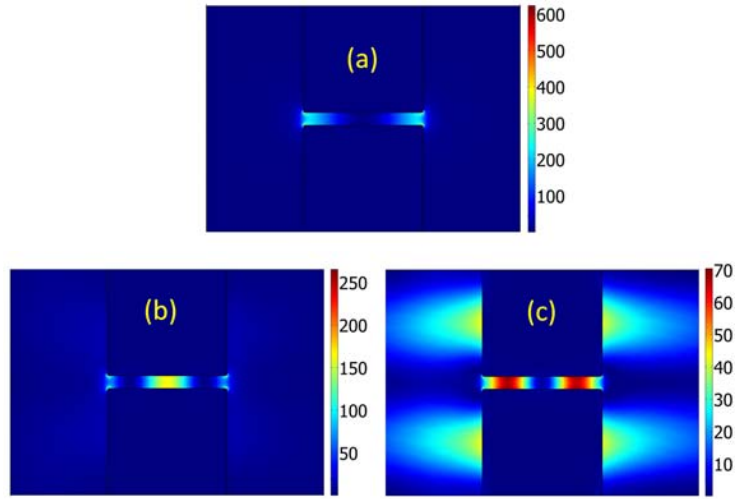

**Fig.4:** Total electric field intensities for the three pumping (solid green arrows) configurations shown in Fig.3. The local intensity is largest when the $\lambda_p \sim$1115nm, at the Fabry-Perot (FP) resonance (a). The resonance to the right of the plasmonic band gap, 625nm, (b) is reminiscent of a FP mode. The resonance to the left of the gap, 575nm, (c) is a hybrid mode that simultaneously displays a surface wave with the characteristic shape of plasmonic excitation and a cavity mode [35].

We now focus on configuration ($AS_1,P_1,S_1$) shown in Fig.3. All fields are tuned within the Fabry-Perot resonance. In Fig.5 we show the predicted Stokes ($\lambda_S$=1130nm) and anti-Stokes ($\lambda_{AS}$=1100nm) spectra for a single shot, along with the unscaled spectrum of a 240fs pump pulse tuned to $\lambda_p$=1115nm.

Ensemble averages obtained by combining dozens of shots may yield slightly different results. Incident pulses of narrower bandwidth do not modify the relative conversion efficiencies because the resonance is resolved with a pulse only a few femtoseconds in duration. However, longer pulses can resolve Raman lines that may be closely spaced relative to each other. As illustrated in Fig.1, the nanocavity is filled with active material and the external layer that coats the metal is about 4nm thick. In addition to the broad FP bandwidth discernible in Fig.3, another advantage that this type of grating offers is that

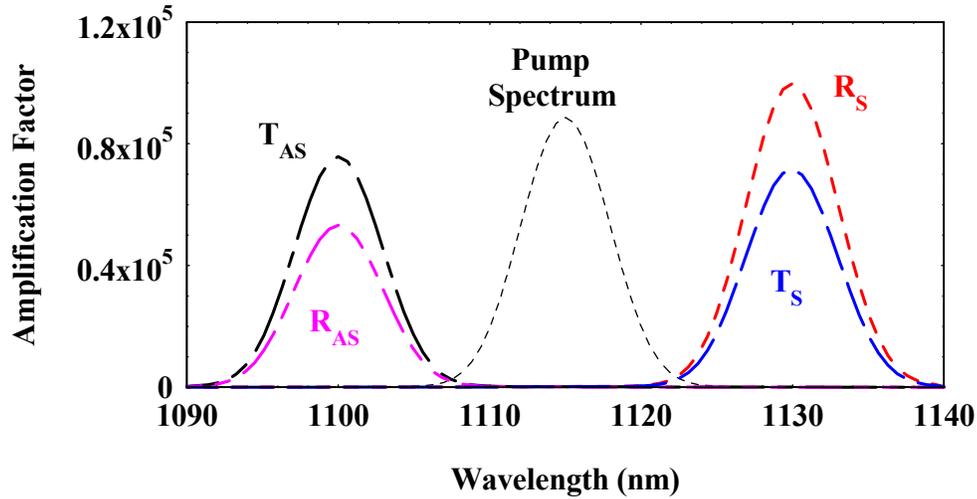

**Fig.5:** Incident pump spectrum for a ~240fs pulse tuned to 1115nm, and typical transmitted (label T) and reflected (label R) generated Stokes (subscript S), and anti-Stokes (subscript AS) spectra, centered at 1130nm and 1100nm, respectively. The amplification factor is determined with respect to a layer of active material 4nm thick, in the absence of metal.

signals are generated in both directions with comparable amplitudes. In this case Stokes and anti-Stokes signals are amplified by a factor of $\sim 10^5$ relative to the case when the metal nanostructure is absent. A study of other tuning conditions reveals that the best scenarios for conversion efficiency (defined as the energy in the generated field normalized by the incident pump energy) arise when the pump is tuned to the FP resonance: this tuning condition provides extended modal volume accompanied by the largest available local fields

In Fig.6(a-c) we show the predicted spectra of the generated anti-Stokes signals for tuning conditions that match both plasmonic band edges (575nm and 625nm) and the FP resonance with near-zero anti-Stokes shift (~1115nm). The figure's inset shows that the normalized gain, or enhancement factor, calculated using Eq.(21) predicts reasonably well the relative conversion efficiencies at the plasmonic band edges. In contrast, both Eqs.(21-22) fall short for tuning conditions near the FP resonance and fail entirely to discriminate between transmitted and reflected amplitudes. However, a comparison of Figs.6(a-c) and Fig.3 shows that for this type of nanostructure there is a rather suggestive and striking correlation between *linear* absorption (proportional to $\mathrm{Im}(\varepsilon)|E_{\omega_p}^{local}|^2$) and Raman gain.

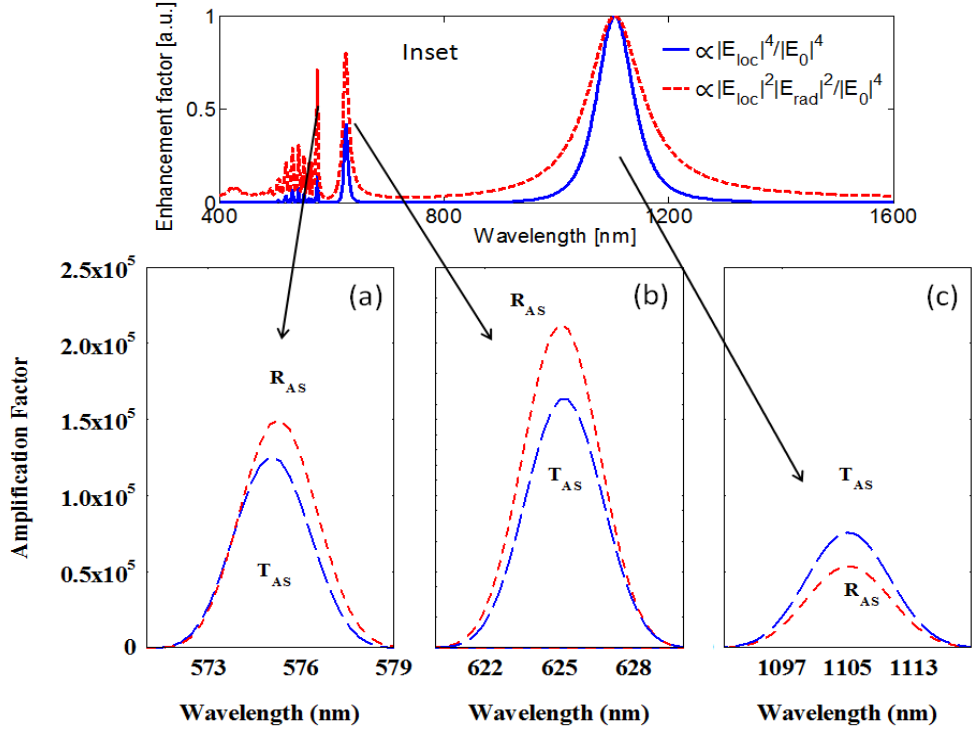

**Fig.6:** Transmitted ($T_{AS}$) and reflected ($R_{AS}$) anti-Stokes (AS) spectra obtained by tuning the AS field to (a) 575nm, (b) 625nm, and (c) 1105nm. 120fs pump pulses are tuned to 1115nm. **Inset**: normalized Raman enhancement factors, according to Eq.(21) (dashed red curve) , and Eq.(22) (solid blue curve). Comparing Fig.6(a), (b) and (c) and the inset shows that Eq.(21) predicts fairly well the relative conversion efficiencies near the plasmonic band gap. Both Eqs.(21-22) fail near the FP resonance, while neither gives insight regarding the proportions of forward and backward generation.

**Grooved Grating**

In Fig.7 we show the linear reflection and absorption spectra for the grating of Fig.1(b), obtained at normal incidence with *p*=1030nm, along with three possible tuning configurations. The reflection (absorption) minima (maxima) observable above 350nm are associated with energy transferred to the substrate in the form of surface waves, accompanied by local field maxima. On the other hand, the minimum near 320nm is related to the plasma frequency of silver. The tuning condition that ostensibly yields the largest conversion efficiency and a reflection amplification factor of $\sim 4\times 10^5$ (see Fig.8) in Fig.7 is configuration ($AS_3,P_3,S_3$), with $\lambda_p$=1066nm and $\lambda_{AS}$=545nm. The results in Figs.8(a-b) show that tuning the anti-Stokes field at 545nm generates nearly ten times more photons compared to the case of near zero anti-Stokes shift. This is in sharp contrast with the predictions obtained using the Raman enhancement factors defined in Eqs.(21-22), which yield maximum efficiency for zero Stokes shift. In addition, our calculations show that for a grating thickness $t$=200nm photons generated either at $\lambda_{AS}$=545nm or near $\lambda_{AS}$=1066nm can tunnel through the structure and propagate in the forward direction at a rate that is still nearly two orders of magnitude larger compared to the case when the grating is absent. Similar results hold for the Stokes signal.

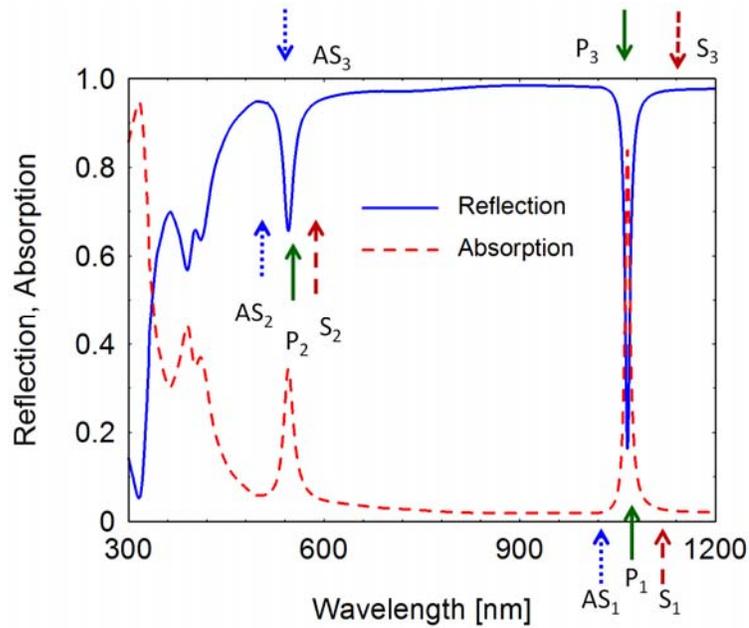

**Fig.7:** Reflection and absorption spectra vs. incident wavelength for the grating in Fig.1(b); $a$=300nm, $w$=45nm, $t$=200nm, and $p$=1030nm. The solid green, dashed red and dotted blue arrows indicate pump, Stokes, and anti-Stokes tuning conditions, respectively, for three possible configurations. Reflection (absorption) minima (maxima) above 350nm correspond to surface wave excitations. In [Media2](Media2) we show the fields for tuning conditions ($AS_3,P_3,S_3$), for a 350fs pump pulse. The actual position of the $S_3$ wave is at a wavelength longer than what is displayed in the figure.

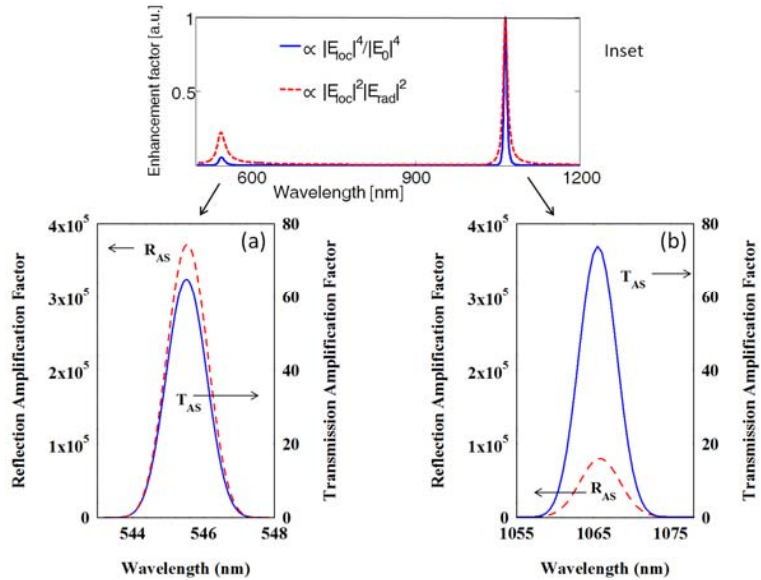

**Fig.8:** Transmitted ($T_{AS}$) and reflected ($R_{AS}$) anti-Stokes (AS) spectra obtained by tuning 350fs pump pulses at 1066nm, and the AS field in turn to 545nm (a), and 1065nm (b). We note that the in both cases the transmitted signal is far smaller compared to the reflected component, but is still nearly two orders of magnitude more intense compared to the case when the metal grating is absent, as the generated signal tunnels through the metal. **Inset**: Raman enhancement factors $G$ (dashed, red curve) and $G_s$ (solid, blue curve).

**Metal Nanowire on Metal Substrate**

In Fig.9(a) we show linear reflection and absorption for the periodic structure depicted in Fig.1(c), which consists of a silver nanowire having radius $r$=36nm placed a distance $d$=0.6nm away from a silver substrate 50nm thick, with periodicity $p$=240nm. Near the resonance at 600nm, the linear transmission is below 1%, similar to reflection, but we do not show it for clarity. We note that grating periodicity here plays a minor role because we are operating at wavelengths well-above $p$. The excited field modes are then related to localized resonances on the surface of the nanowire, modified by the presence of the substrate [43]. In Figs.9(b) and (c) we plot snapshots of the local electric (Fig.9(a)) and magnetic field (Fig.9(b)) intensities normalized with respect to the incident field intensity when a 50fs pulse excites the system. The local field intensity is enhanced more than four orders of magnitude with respect to the incident intensity at the absorption maximum (reflection minimum) near 600nm, and increases extremely rapidly as $d \Rightarrow 0$, with enhancement factors that can easily exceed ten orders of magnitude as sharp metal tips form at the nanowire/substrate contact point. In Fig.10 we show that for silver the predicted Stokes (~610nm) and anti-Stokes (~590nm) amplification factors exceed $10^7$ for the reflected fields, while both transmitted signals are amplified by factors larger than $10^5$.

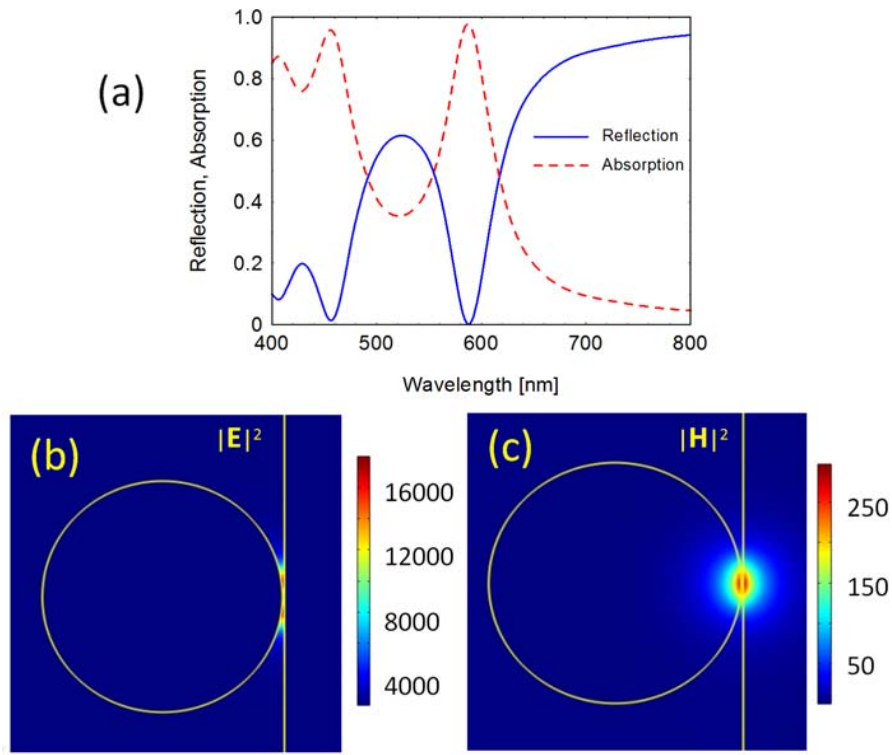

**Fig.9:** (a) Linear reflection and absorption spectra for the grating depicted in Fig.1(c). The bandwidth of the resonance near 600nm is ~60nm wide, and can easily accommodate an ultrashort pulse. (b) Snapshot of the total local electric and (c) magnetic field intensities excited by an incident 50fs pulse tuned at absorption peak at ~600nm. The local electric field intensity is enhanced by four orders of magnitude, and manifests singular behavior when the nanowire touches the substrate.

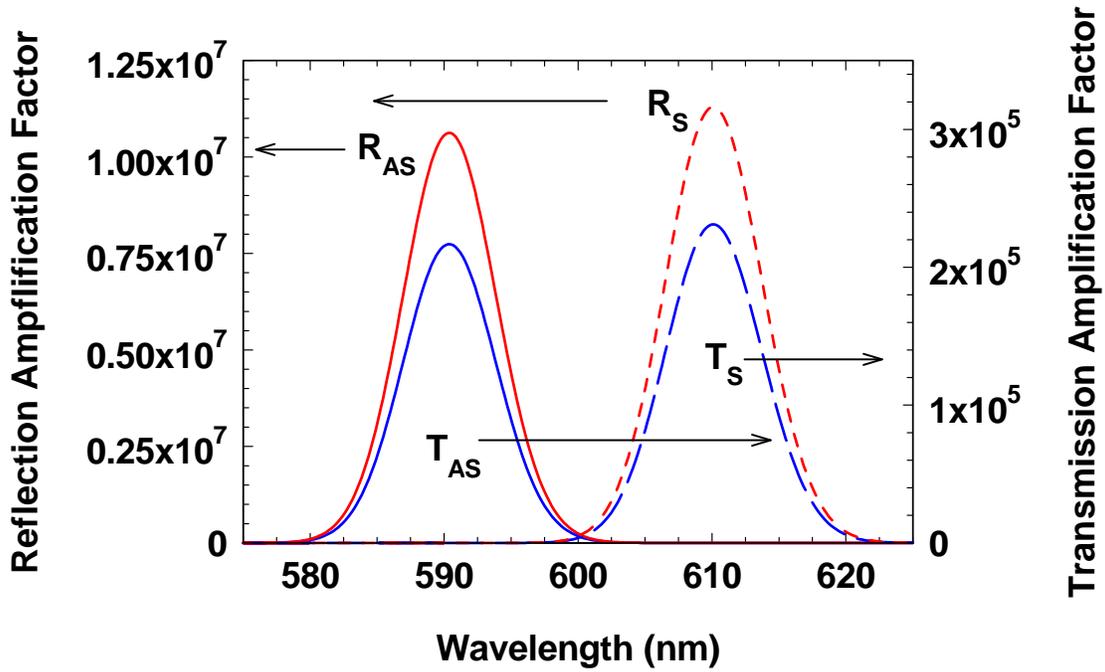

**Fig.10:** Transmitted (T, blue curves, right axis) and reflected (R, red curves, left axis) Stokes (S) and anti-Stokes (AS) spectra obtained by tuning 50fs pump pulses to 600nm, and S and AS fields tuned to 590nm and 610nm, as indicated.

**Local Field Effects and Raman Gain Saturation**

Even though our system displays unexceptionally high local field intensity enhancement factors (~$10^4$-$10^5$), the last example shows that slight geometrical modifications, like placing the nanowires closer to the substrate, have the potential to push local fields enhancement up by well over ten orders of magnitude, with Raman amplification factors that may be much larger than our predicted $10^7$ (at least in principle, neglecting for a moment fabrication issues). The first question that comes to mind is this: How does the atomic system respond in the limit as $d \Rightarrow 0$? Near sharp tips the fields tend to become singular and stimulated emission is sure to dominate, raising doubts about the assumptions that within the hot spots the medium always remains in the ground state; whether or not saturation phenomena are triggered; or if the breakdown intensity of the medium is reached. We now briefly explore this limit.

Our model consists of a full-wave approach where local field effects are included because we preserve the spatial dependence of the fields and allow local dipoles to interact. As a result, there is no distinction between the local and the Maxwell fields at every point in space. In order to have a sense of how the local field contributes to the process we simplify the system and assume that the dipole moment has only one component. In turn this selects one field constituent, say the longitudinal portion. Removing any vector notation, the local field $E_L$ and the Maxwell field $E$ are related as follows [57]:

$$E_L = E + \frac{4\pi}{3} P, \qquad (23)$$

where the polarization $P$ is given by the scalar form of Eq.(19). Retaining the assumption the system remains in the ground state ($W \approx -1$), Eq.(14) becomes:

$$\dot{Q} = -\Gamma Q + \frac{\mu_0^2}{\hbar^2 \omega_{01}} E_L^2 e^{i\delta\tau}, \qquad (24)$$

where $\Gamma$ is now a generalized, complex damping coefficient. Assuming the presence of one Stokes and one anti-Stokes fields, with the aid of Eqs.(5) and (19) we square the local field of Eq.(23) and substitute lowest order terms into Eq.(24) to obtain:

$$\dot{Q} = \left(-\Gamma + i\frac{\mu_0^2}{\hbar^2\omega_{01}}\frac{8\pi\chi_L}{3}|E_{Total}|^2\right)Q + \frac{\mu_0^2}{\hbar^2\omega_{01}}\left(\mathcal{E}_P\mathcal{E}_S^* + \mathcal{E}_{AS}\mathcal{E}_P^*\right)\left(1 + \frac{8\pi\chi_L}{3} - i\frac{16\pi^2\chi_L^2}{9}|Q|^2\right), \qquad (25)$$

where we have defined the total field intensity $|E_{Total}|^2 = |\mathcal{E}_P|^2 + |\mathcal{E}_S|^2 + |\mathcal{E}_{AS}|^2$, and $\mathcal{E}_P, \mathcal{E}_S$ and $\mathcal{E}_{AS}$ are pump, Stokes and anti-Stokes field amplitudes. In Eq.(25) we recognize a dynamic, intensity-dependent detuning as well as additional linear and nonlinear terms. We also note that on the right hand side of the equation the nonlinear term proportional to $|Q|^2$ may be neglected *only* in gases, where $\chi_L \ll 1$. Confining ourselves to low-density media, if the intensity were high enough the time dependence of the atomic coherence could be neglected and an approximate solution for $Q$ written as:

$$Q \approx i\frac{3}{8\pi\chi_L}\frac{\left(\mathcal{E}_P\mathcal{E}_S^* + \mathcal{E}_{AS}\mathcal{E}_P^*\right)}{\left(|\mathcal{E}_P|^2 + |\mathcal{E}_S|^2 + |\mathcal{E}_{AS}|^2\right)}. \qquad (26)$$

Substituting Eq.(26) into Eqs.(20), the assumption that the pump remains undepleted and the generated fields are small leads to the following dependence of the nonlinear Stokes and anti-Stokes sources:

$$\begin{aligned}\mathcal{P}_{\omega_S}^{NL} &\propto \left(\mathcal{E}_S + \mathcal{E}_{AS}^*\right) \\ \mathcal{P}_{\omega_{AS}}^{NL} &\propto \left(\mathcal{E}_S^* + \mathcal{E}_{AS}\right)\end{aligned} \qquad (27)$$

Thus, via Eqs.(26-27) our model predicts that at very high intensities local field effects introduce an intensity-dependent detuning that linearizes and saturates Stokes and anti-Stokes gain. In any case, under extreme intensity conditions an assessment of molecular inversion is probably in order, and may be accomplished by including the combined dynamics of Eqs.(16-17).

**Conclusions**

In summary, we have derived a semi-classical model of surface enhanced stimulated Raman scattering by assuming a quantized Raman active medium, and Stokes and anti-Stokes signals that are generated from quantum noise. Using the model we have predicted Stokes and anti-Stokes amplification factors between $10^5$ and $10^7$ for a variety of metal-based nanostructures that range from the highly transmissive and operate in the enhanced transmission regime, to the opaque, such as grooved gratings and metal nanowires placed near metal substrates. We have shown that the Raman enhancement factors defined in Eqs.(21-22), derived from the computation of linear fields, can be misleading in that they may fail to provide accurate information about relative gain between Stokes or anti-Stokes lines, cannot discriminate between reflected and transmitted fields, and may be at odds with each other, depending relative on Stokes and anti-Stokes spectral positions. We have also discussed the limits of validity of our theory, shown that in the reduced two-level atom view the medium may become inverted within hot spots, and demonstrated that local field effects can help reign in and quench Raman gain in the high-intensity regime.